\documentclass[useAMS,usenatbib]{mn2e}
\usepackage{psfrag,graphicx}
\usepackage{color}
\usepackage{txfonts}
\usepackage{breqn}
\title[High resolution studies of massive primordial haloes]
  {High resolution studies of massive primordial haloes}
\author[M.~A.~Latif et al.]
  {M.~A.~Latif,$^1$
  D.~R.~G.~Schleicher,$^1$ 
  W.~Schmidt,$^1$
  J.~Niemeyer$^1$
   \newauthor % starts a new line in the
   $^1$ Institut f\"ur Astrophysik, Georg-August-Universit\"at, \\
    Friedrich-Hund-Platz 1, D-37077 G\"ottingen, Germany}
\date{today}

\pagerange{\pageref{firstpage}--\pageref{lastpage}} \pubyear{2009}

\def\LaTeX{L\kern-.36em\raise.3ex\hbox{a}\kern-.15em
      T\kern-.1667em\lower.7ex\hbox{E}\kern-.125emX}

\begin{document}

\bibliographystyle{mn2e}

\label{firstpage}

 \maketitle
 \begin{abstract}
{Atomic cooling haloes with virial temperatures $\rm T_{vir} \geq 10^{4}$ K are the most plausible sites for the formation of the first galaxies and the first intermediate mass black holes. It is therefore important to assess whether one can obtain robust results concerning their main properties from numerical simulations. A major uncertainty is the presence of turbulence, which is barely resolved in cosmological simulations. We explore the latter both by pursuing high-resolution simulations with up to $64$ cells per Jeans length and by incorporating a subgrid-scale turbulence model to account for turbulent pressure and viscosity on unresolved scales. We find that the main physical quantities in the halo, in particular the density, temperature and energy density profile, are approximately converged. However, the morphologies in the central $500$~AU change significantly with increasing resolution and appear considerably more turbulent. In a systematic comparison of three different haloes, we further found 
that the turbulence subgrid-scale model gives rise to more compact central structures, and decreases the amount of vorticity.  Such compact morphologies may in particular favor the accretion onto the central object.
%
%
%. In this article, we aim to study the implications of gravity driven turbulence in protogalactic haloes. By varying the resolution per Jeans length, we explore whether the turbulent cascade is resolved well enough to obtain converged results. We have performed high resolution cosmological simulations using the adaptive mesh refinement code Enzo including a subgrid-scale turbulence model to study the role of unresolved turbulence. We compared the results of three different Jeans resolutions from 16 to 64 cells.
%While radially averaged profiles roughly agree at different resolutions, differences in the morphology reveal that even the highest resolution employed provides no convergence. Moreover, taking into account  unresolved turbulence  significantly influences the morphology of a halo. We have quantified the properties of the high-density clumps in the halo. These clumps are gravitationally unbound with temperature above 6000 K and densities of the order of $\rm 10^{-12}$ g~cm$^{-3}$. In general, the clumps with SGS turbulence are denser and more massive compared with their counterparts in the standard simulation setup that ignores unresolved turbulence.  
}

 \end{abstract}


\section{Introduction}

The nonlinear interplay of gravity, fluid dynamics, and radiative cooling lies at the heart of the continuing challenge to predict the properties of the first astrophysical objects of the universe. Scale separation, which is one of the most important prerequisites for robust analytic or numerical calculations, does not apply: whereas the kinetic energy budget is dominated by large-scale modes, fragmentation and cooling instabilities grow fastest on those scales with the strongest density fluctuations, i.e., those close to the Jeans scale. In order to capture the relevant degrees of freedom, grid-based hydrodynamical simulations employ adaptive mesh refinement over many orders-of-magnitude. 

In addition to the masses and multiplicity of the first stars, the question of whether and when supermassive black holes (SMBH) can form by direct collapse of a primordial gas cloud belongs to the key topics of current research \citep{2002ApJ...569..558O,2003ApJ...596...34B,2006ApJ...652..902S,2006MNRAS.370..289B,2008MNRAS.391.1961D,2008arXiv0803.2862D,2010MNRAS.402.1249S,2010ApJ...712L..69S,2011MNRAS.411.1659L}. Massive primordial haloes assembled at redshifts 10-15 are the plausible sites to host the first galaxies formed at the end of cosmic dark ages. They are also potential candidates for the formation of intermediate mass black holes through direct collapse \citep{1984ARA&A..22..471R, 2003ApJ...596...34B, 2004MNRAS.354..292K, 2006MNRAS.370..289B, 2009ApJ...702L...5B, 2010A&ARv..18..279V}. Thus, their understanding is a matter of great astrophysical interest. 
%In the simulations presented below, we considered high-mass halos ($\rm M > 10^{6}~M_\odot$). Our prime objective is to study the effect of resolution on the global properties and the morphology of atomic cooling haloes. We intend to focus on the direct collapse scenario for SMBH in future work.

The first stars, on the other hand, are expected to form in so-called minihalos with $\rm 10^5-10^6$~M$_\odot$ \citep{2002Sci...295...93A, 2004PASP..116..103B, Yoshida08}. In a recent study, \citet{2012ApJ...745..154T} found that an increased resolution per Jeans length results in non-convergence of global properties. They discovered that enhanced Jeans resolution produces higher infall velocities, increased temperatures as well as decreased content of molecular hydrogen. As convergence has not been achieved even at the highest resolutions, major uncertainties are present concerning the expected accretion rates, temperature distributions and the fragmentation behavior. It is therefore important to assess whether similar restrictions apply to other systems, in particular the so-called atomic cooling haloes. In the study presented here, we therefore assess the convergence behavior employing numerical simulations with a resolution of $\rm 16$, $\rm 32$ and $\rm 64$ cells per Jeans length. We focus on halos 
exposed to 
strong Lyman Werner radiation, as these are the candidates for black hole formation via direct collapse. We note that these are the highest resolution studies to date, with previous studies typically employing a resolution of $\rm 16$ cells per Jeans length \citep{2008ApJ...682..745W, 2009MNRAS.393..858R, 2010MNRAS.402.1249S}.

The need for high-resolution investigations has been previously derived in different contexts. In particular, \citet{2011ApJ...731...62F} reported that the turbulent energy as in collapsing gas clouds converges only for a resolution of at least $\rm 32$ cells per Jeans length. The turbulent energy in these clouds is released from the gravitational potential, due to the initial deviations from spherical symmetry \citep{1953ApJ...118..513H,1982ApJ...258L..29S,2010A&A...520A..17K,2010ApJ...712..294E}.

It is well known from the field of contemporary star formation that turbulence has important implications for the density PDF, clump statistics, and angular momentum transport in gravitationally unstable gas clouds (e.g., \citep{1981MNRAS.194..809L,2004RvMP...76..125M,2007ARA&A..45..565M,2012arXiv1209.2856F}). During high-redshift structure formation, turbulence was shown to play a major role in so-called minihalos. In particular, it regulates the angular momentum transport and delays the formation of a disk \citep{2002Sci...295...93A, 2004PASP..116..103B, Yoshida08}, but also influences the fragmentation behavior \citep{Clark11, Smith11, Greif12}. In more massive halos, the presence of turbulence was reported by \cite{2008MNRAS.387.1021G} and \cite{2008ApJ...682..745W}. While the implications were not explored in detail, it is expected that it can influence the formation of intermediate mass black holes through its impact on disk formation and the angular momentum distribution. We note that the presence 
of disks 
is a central assumption in direct collapse models \citep{2004MNRAS.354..292K, 2006MNRAS.370..289B,2007NCimR..30..293L,2009ApJ...702L...5B}, and an efficient means of angular momentum transport is generally required to allow the formation of a massive central object. \citet{2009ApJ...702L...5B} therefore invoked the presence of nested bar-like instabilities to provide sufficient means for angular momentum transport. While our simulations do not exactly support this generation mechanism, they show that turbulence is efficiently produced during gravitational collapse.
% Previous studies exploring the turbulence in collapsing gas clouds found that the resolution per Jeans length is important to obtain converged turbulent energies \citep{2011ApJ...731...62F}. Similar results are reported by \cite{2012ApJ...745..154T} for the minihaloes. 

Such turbulence will not only influence angular momentum transport, but also amplify existing weak magnetic fields via the small scale dynamo \citep{1968JETP...26.1031K,1998MNRAS.294..718S,2005PhR...417....1B,Schobera,2010A&A...522A.115S,2010ApJ...721L.134S,2011ApJ...731...62F, 2012ApJ...745..154T,Schoberb}. The enhanced magnetic field strength could exert an additional pressure, and further contribute to the angular momentum transport. More importantly, this initial growth via the small-scale dynamo provides a strong initial field on which the $\rm \alpha-\Omega$ dynamo can act to produce the coherent fields observed today \citep{Beck96, Arshakian09}.

We note that an accurate modeling of turbulence not only requires high numerical resolution, but also a consistent treatment of the unresolved scales. For this purpose, we explore the implications of a turbulence subgrid-scale (SGS) model for the turbulent kinetic energy \citep{2009ApJ...707...40M,2011MNRAS.414.2297I,2011A&A...528A.106S}. Its main features are non-ideal terms in the fluid equations produced by an effective turbulent viscosity, an effective turbulent pressure, and fully time and space dependent dissipation of turbulent kinetic energy into thermal energy. All of these terms make order unity contributions for transonic flows, which is the case here. In this study, we therefore employ high-resolution studies of massive halos to explore their central properties. A central question is in particular for which properties convergence can be achieved, which is directly relevant for the direct collapse scenario concerning the formation of intermediate mass black holes.

Our paper is organized as follows. In the next section, we describe the simulation setup and the numerical methods employed. In the 3rd section of the paper, we present the results obtained in this study. In the last section of this article, we summarize our main results and confer our conclusions.

\section{Numerical Methods and simulation details}

A modified version of the Enzo code including the subgrid-scale model for unresolved turbulent fluctuations (see below for a description) has been used to perform the simulations presented in this work. Enzo is an adaptive mesh refinement (AMR), parallel, grid-based cosmological hydrodynamics code \citep{2004astro.ph..3044O,2007arXiv0705.1556N}. It can run on massively parallel systems and has been used for a  wide variety of astrophysical applications. The message passing interface (MPI) is used to achieve portability and scalability on different systems. The computational domain is discretized into nested grid cells. It has two exchangeable grids, a uniform grid and a block-structured adaptive grid. We use a split hydro solver with a 3rd order piece-wise parabolic (PPM) method for hydrodynamical calculations. The dark matter N-body dynamics is solved using the particle-mesh technique. A multigrid Poisson solver is employed for the self-gravity computations. 
  
\subsection{Initial conditions}

The simulations are started with the cosmological initial conditions generated from Gaussian random fields. We employ the inits package available with the public version of the Enzo code to create nested grid initial conditions. Our simulations start at redshift $\rm z=99$ with a top grid resolution of $\rm 128^{3}$ cells and we select the massive halo at redshift 15 using the halo finder of \cite{2011ApJS..192....9T}. Two initial nested levels of refinement are subsequently added each with a resolution of $\rm 128^{3}$ cells. Our simulation box has a cosmological size of 1 Mpc $\rm h^{-1}$ and is centered on the massive halo. In total, we initialize 6291456
particles to compute the evolution of the dark matter dynamics and have a final dark matter resolution of 300 $\rm M_{\Theta}$. While our dark matter halo is thus well-resolved, we note that additional fluctuations could be present in case of a higher resolution in dark matter. The parameters for creating the initial conditions and the distribution of baryonic and dark matter components are taken from the WMAP seven years data \citep{2011ApJS..192...14J}. We further allow additional 27 levels of refinement in the central 62 kpc region of the halo during the course of simulation. It gives us a total effective resolution of 3 AU in comoving units. The resolution criteria used in these simulations are based on the Jeans length, the gas over-density and the particle mass resolution. The grid cells matching these requirement are marked for a refinement.

The simulations conducted in this work mandated the Jeans length resolution of 16, 32 and 64 cells throughout their evolution. This criterion was applied during the course of simulations to ensure that all physical processes like shock waves and Truelove criterion \citep{1997ApJ...489L.179T} are well resolved. We stop the simulations after they reach the maximum refinement level and start to violate the Jeans criterion. The results at later stages would not be reliable.  

\subsection{Chemistry}

To include the primordial non-equilibrium chemistry, the rate equations of the following 9 species: $\rm H,~H^{+},~He,~He^{+},~He^{++},~e^{-},~H^{-},~H_{2},~H_{2}^{+}$ are self-consistently solved in the cosmological simulations. We make use of $\rm H_{2}$ photo-dissociating background UV flux implemented in the Enzo code. An external UV field of constant strength $\rm 10^{3}$ in units of $\rm J_{21}$ is used in the simulations. We presume that such flux is generated from a nearby star forming halo \citep{2008MNRAS.391.1961D} and is emitted by Pop III stars with a thermal spectrum of $\rm 10^{5}$ K. We include several cooling and heating mechanisms like collisional ionization cooling, radiative recombination cooling, collisional excitation cooling, H$_{2}$ cooling as well as $\rm H_{2}$ formation heating. The chemistry solver used in this work is a modified version of \cite{1997NewA....2..181A,1997NewA....2..209A}. 

We note that for columns above $\rm 10^{22}~ cm^{-2}$, the gas becomes optically thick to Lyman $\alpha$ photons, providing a potential complication for the further evolution. In fact, \citet{2006ApJ...652..902S} suggested that the latter may effectively stop the cooling and provide a transition to an approximately adiabatic regime. The latter was explored in detail by \citet{2010ApJ...712L..69S}, finding that at this point, additional processes become relevant, including the two-photon decay (2s-1s transition) and H$^-$ formation cooling. Effectively, the temperature evolution is then very close to the evolution obtained from optically thin Lyman $\alpha$ cooling. We also note that for a stellar spectrum of $\rm 10^{5}$ K $\rm H_{2}$ is mainly dissociated by the Solomon process while for the stellar spectrum of $\rm 10^{4}$ K the main dissociation route is $\rm H^{-}$ \citep{2010MNRAS.402.1249S,2011MNRAS.418..838W}. In principle, one would of course expect the presence of both contributions. 
The details of these processes however would not matter as long as H$_2$ is efficiently dissociated.

\subsection{SGS turbulence Model}
Due to the high Reynolds numbers relevant for astrophysical systems, it is not possible to resolve all scales down to the dissipative scale even with adaptive mesh refinement techniques. Turbulence cascades from coarser grids corresponding to large scales down to the center of the structure forming haloes without being properly accounted for. In engineering and many other disciplines of computational fluid dynamics subgrid scale turbulence models are used to represent the effect of unresolved turbulence on resolved scales. The unresolved turbulence has gained lot of interest in astrophysical simulations \citep{2008ApJ...686..927S,2009MNRAS.395.1875O,2009ApJ...707...40M}. To compute the unresolved turbulence on grid scales in our simulation, we use the subgrid scale (SGS) turbulence model by \citet{SchmNie06b}. This SGS model is based on a mathematically rigorous approach separating  the resolved and unresolved scales, and connecting them via an eddy-viscosity closure for the non-linear energy transfer 
across the grid scale. The turbulent viscosity is given by the grid scale and the SGS turbulence energy,
i.~e., the kinetic energy associated with numerically unresolved turbulent velocity fluctuations. The equations for compressible fluid dynamics are decomposed into resolved (large scales) and unresolved (small scales) parts using the filter formalism proposed by \citet{1992JFM...238..325G} in terms of density weighted quantities.
% \ch{Any field quantity $a$ is divided into a smoothed part $\rm \langle a \rangle $ and a fluctuating part $\rm a'$, where $\rm \langle a \rangle $ varies only at scales greater than the filter length. The density weighted filtered quantities are defined as
% \ch{\begin{equation}
% \hat{a} =  {\langle \rho a  \rangle \over \langle \rho   \rangle},
% \label{eq1}
% \end{equation}}
% and the generalized moments are denoted by:}
% \ch{\begin{equation}
% \hat{\tau}(a,b) =  \langle \rho a b \rangle -\langle \rho \rangle \hat{a}\hat{b}
% \label{eq2}
% \end{equation}}
Applying the filtering mechanism to the fluid equations and solving them in comoving coordinate system, we obtain \citep{SchmNie06b,2009ApJ...707...40M}:
% \ch{\begin{equation}
% \begin{eqnarray*}
% {\partial \over \partial t} \langle \tilde{\rho} \rangle + {1 \over a } { \partial \over \partial x_{j}} \hat{u_{j}} \langle \tilde{\rho} \rangle = 0  \\
% { \partial \over \partial t} \langle \tilde{\rho} \rangle \hat{u_{j}} + {1 \over a }{\partial \over \partial x_{i}} \hat{u_{j}} \langle \tilde{\rho} \rangle \hat{u_{i}} = -{1 \over a } {\partial \over \partial x_{j}} \langle \tilde{p} \rangle   +  \langle \tilde{\rho} \rangle \hat{g_{i}^{*}}    \\
%  -{1 \over a }{\partial \over \partial x_{j}}\hat{\tau(u_{i},u_{j})}  -{\dot{a} \over a } \langle \tilde{\rho} \rangle \hat{u_{j}} \\
% { \partial \over \partial t} \langle \tilde{\rho} \rangle e_{res} + {1 \over a }{\partial \over \partial x_{j}} \hat{u_{j}} \langle \tilde{\rho} \rangle e_{res} =  -{1 \over a } {\partial \over \partial x_{i}} \hat{u_{i}} \langle \tilde{p} \rangle  - {\dot{a} \over a } (\langle \tilde{\rho} \rangle e_{res} + {1 \over 3 }  \\
%  \langle \tilde{\rho} \rangle \hat{u_{i}}\hat{u_{i}}  + \langle \tilde{p} \rangle ) + \langle \tilde{\rho} \rangle(\lambda + \epsilon)-{1 \over a } \hat{u_{i}} {\partial \over \partial x_{j}} \hat{\tau(u_{i},u_{j})}  \\
% { \partial \over \partial t} \langle \tilde{\rho} \rangle e_{t} + {1 \over a } { \partial \over \partial x_{j}} \hat{u_{j}} \langle \tilde{\rho} \rangle e_{t} = \mathbb{D} + \Gamma - \langle \rho \rangle (\lambda + \epsilon)  - {1 \over a } \hat{\tau(u_{i},u_{j})} {\partial \over \partial x_{j}} \hat{u_{i}} \\
% - 2{\dot{a} \over a } \langle \tilde{\rho} \rangle e_{t}
% \end{eqnarray*}
\begin{equation}
 {\partial \over \partial t} \langle \tilde{\rho} \rangle + {1 \over a } { \partial \over \partial x_{j}} \hat{u_{j}} \langle \tilde{\rho} \rangle = 0 
\end{equation}
\begin{dmath}
{ \partial \over \partial t} \langle \tilde{\rho} \rangle \hat{u_{j}} + {1 \over a }{\partial \over \partial x_{i}} \hat{u_{j}} \langle \tilde{\rho} \rangle \hat{u_{i}} = -{1 \over a } {\partial \over \partial x_{j}} \langle \tilde{p} \rangle   +  \langle \tilde{\rho} \rangle \hat{g_{i}^{*}} -{1 \over a }{\partial \over \partial x_{j}}\hat{\tau}(u_{i},u_{j}) \\
 -{\dot{a} \over a } \langle \tilde{\rho} \rangle \hat{u_{j}} 
\end{dmath}
\begin{dmath}
 {\partial \over \partial t} \langle \tilde{\rho} \rangle e_{res} + {1 \over a }{\partial \over \partial x_{j}} \hat{u_{j}} \langle \tilde{\rho} \rangle e_{res} =  -{1 \over a } {\partial \over \partial x_{i}} \hat{u_{i}} \langle \tilde{p} \rangle  - {\dot{a} \over a } (\langle \tilde{\rho} \rangle e_{res} + {1 \over 3 }  \langle \tilde{\rho} \rangle \hat{u_{i}}\hat{u_{i}} \\
  + \langle \tilde{p} \rangle ) + \langle \tilde{\rho} \rangle(\lambda + \epsilon)-{1 \over a } \hat{u_{i}} {\partial \over \partial x_{j}} \hat{\tau}(u_{i},u_{j})
\end{dmath}
\begin{dmath}
 { \partial \over \partial t} \langle \tilde{\rho} \rangle e_{t} + {1 \over a } { \partial \over \partial x_{j}} \hat{u_{j}} \langle \tilde{\rho} \rangle e_{t} = \mathbb{D} + \Gamma - \langle \rho \rangle (\lambda + \epsilon)  - {1 \over a } \hat{\tau}(u_{i},u_{j}) {\partial \over \partial x_{j}} \hat{u_{i}} \\
- 2{\dot{a} \over a } \langle \tilde{\rho} \rangle e_{t}
\end{dmath}
here a is the scale factor and the quantities with $\tilde{}$ are the comoving gas density, velocity and pressure etc. $\sigma_{ij}$ is the viscous stress tensor, q is turbulent velocity, $\lambda$ describes the effect of unresolved pressure fluctuations, $\epsilon$ accounts for the dissipation of kinetic energy and $\eta$ is the dynamic viscosity. Equations (1-3) are the mass, the momentum and the energy conservation equations. Equation (4) solves the evolution of SGS turbulent energy ($\rm e_{t}$). The quantities $\mathbb{D}$, $\lambda$, $\epsilon$, $\Gamma$ and $\hat{\tau}(\nu_{i}, \nu_{j})$ are unknowns and are computed in terms of closure relations, i.e., functions of the filtered flow quantities and the turbulent energy. The expressions for the closure terms are the following: 
% \ch{where}
\begin{equation}
\mathbb{D} = { \partial \over \partial r_{i}} C_{\mathbb{D}} \langle \rho \rangle l_{\Delta}q^{2} { \partial \over \partial r_{i}} q
\end{equation}
\begin{equation}
\hat{\tau}_{ij} = -2 \eta_{t} S^{*}_{ij} + {1 \over 3}\delta_{ij} \langle \rho \rangle q^{2} 
\end{equation}
where
\begin{equation}
S^{*}_{ij} = {1 \over 2} \left({\partial \over \partial r_{j}} \hat{\nu_{i}} + {\partial \over \partial r_{i}} \hat{\nu_{j}} \right) + {1 \over 3} \delta_{ij}\langle \rho \rangle q^{2}
\end{equation}
\begin{equation}
\lambda = C_{\lambda}q^{2}{\partial \over \partial r_{i}} \hat{\nu_{i}}
\end{equation}
\begin{equation}
\epsilon = C_{\epsilon} {q^{3} \over l_{\Delta}} \left(1 +\alpha_{1}  M^{2}_{t} \right)
\end{equation}
Further details and numerical validations of the applied closures are given in dedicated studies of \cite{SchmNie06b} and \cite{2009ApJ...707...40M} and are beyond the scope of this article.

For our simulations, the coefficients of  the model are calibrated against subsonic compressible turbulence simulations, comparable to the regime in our simulations \cite[see][]{SchmNie06b}. To apply the SGS model in cosmological AMR simulations, the method of adaptively refined large eddy simulations is employed \citep{2009ApJ...707...40M}. Since SGS turbulence energy depends on the grid scale, which varies on adaptive meshes, energy must be exchanged with the numerically resolved velocity field when the grid is refined or de-refined. This is achieved by assuming the Kolmogorov two-thirds law for the scaling of the turbulent velocity fluctuations. As long as the turbulence is subsonic and nearly isotropic locally, this is a reasonable assumption. In contrast, the method used by \citet{2011ApJ...733...88G} does not calculate the turbulence 
energy on the grid scale, but the energy associated with a characteristic length scale of buoyant bubbles. Our SGS model completely neglects gravity on unresolved length scales and assumes the turbulent cascade from larger resolved scales as the dominant source of unresolved turbulence \citep[for the effects of buoyancy on subgrid scales, see also][]{SchmNie06c}. Of course, this requires that turbulence is sufficiently resolved. The different resolutions considered in this study allow us to estimate biases from  scale separation between the production of turbulence by gravity and the grid scale. 
% <<< WS

\begin{figure*}
\hspace{-4.0cm}
\centering
\begin{tabular}{c}
\begin{minipage}{12cm}
\includegraphics[scale=0.8]{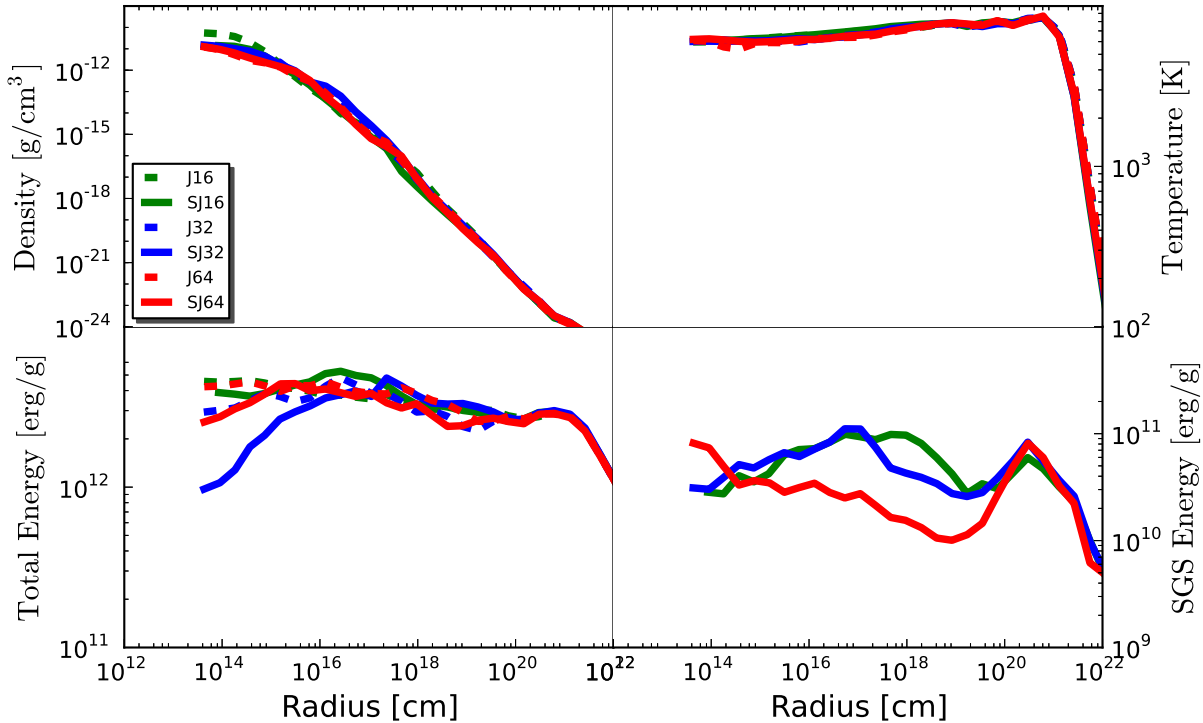}
\end{minipage}
\end{tabular}
\caption{This figure shows the radially binned spherically averaged radial profiles for the halo C with and without SGS turbulence for different Jeans resolutions. The dashed lines show normal runs while the solid lines represent the cases with SGS turbulence as depicted in the legend. The upper left panel of the figure shows the density radial profiles. The temperature radial profiles are depicted the upper right panel. The bottom left panel of the figure shows the averaged total energy radial profiles. The averaged radial profiles of SGS energy are shown in the bottom right panel of the figure.}
\label{fig}
\end{figure*}
% \begin{equation}
% \rm L_{\alpha} =\int dV J(r)
% \label{lum}
% \end{equation}
\begin{figure*}
\hspace{-4.0cm}
\centering
\begin{tabular}{c}
\begin{minipage}{8cm}
\includegraphics[scale=0.6]{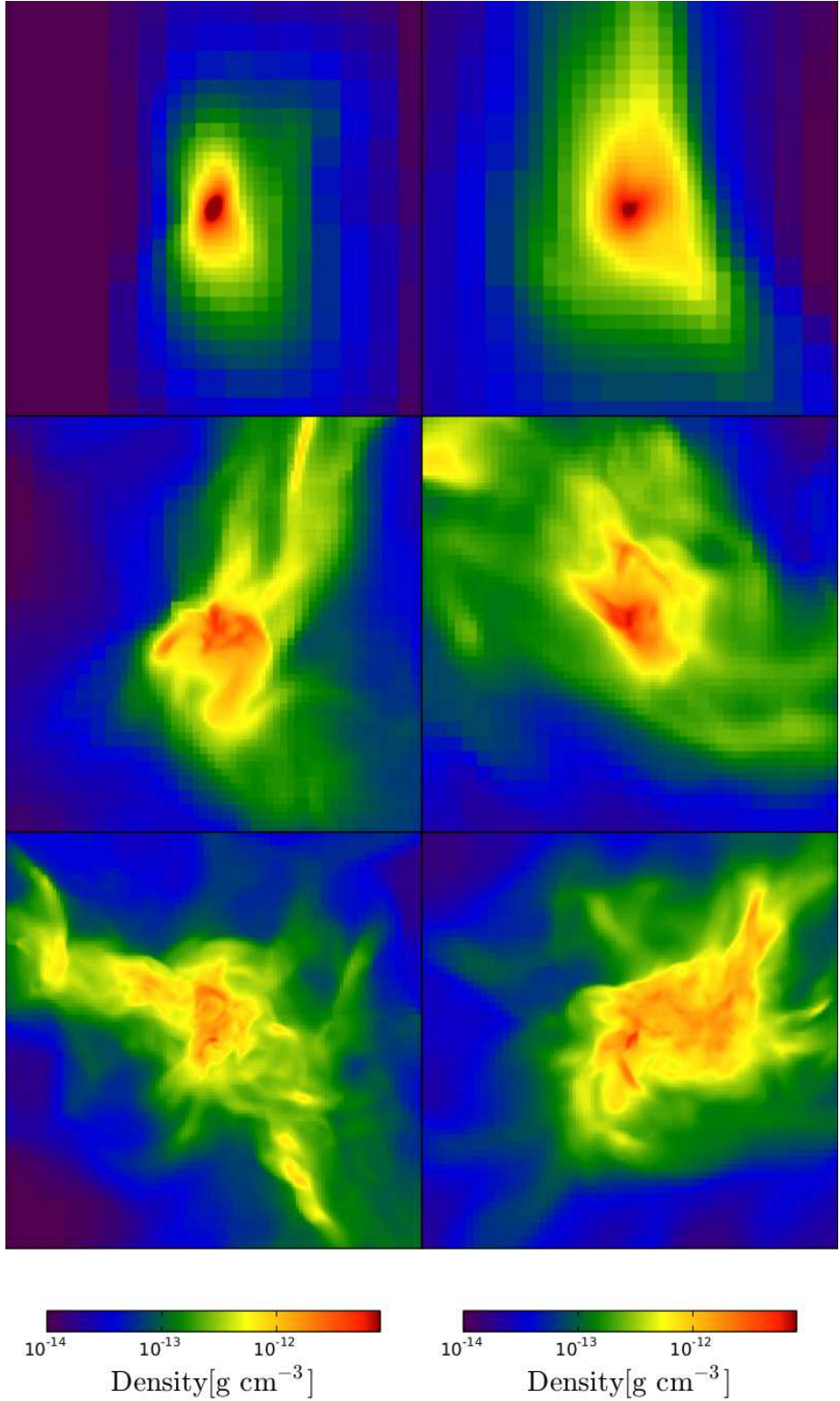}
\end{minipage}
\end{tabular}
\caption{The figure shows the density projections for the halo C with and without SGS turbulence. The state of simulations for the central 500 AU region of the halo is illustrated for various Jeans resolutions. The panels from top to bottom represent the Jeans resolution of 16, 32 and 64 cells respectively. The left panels present the normal runs while the cases with SGS turbulence are shown in the right panels.}
\label{fig0}
\end{figure*}

\begin{figure*}
\hspace{-4.0cm}
\centering
\begin{tabular}{c}
\begin{minipage}{12cm}
\includegraphics[scale=0.8]{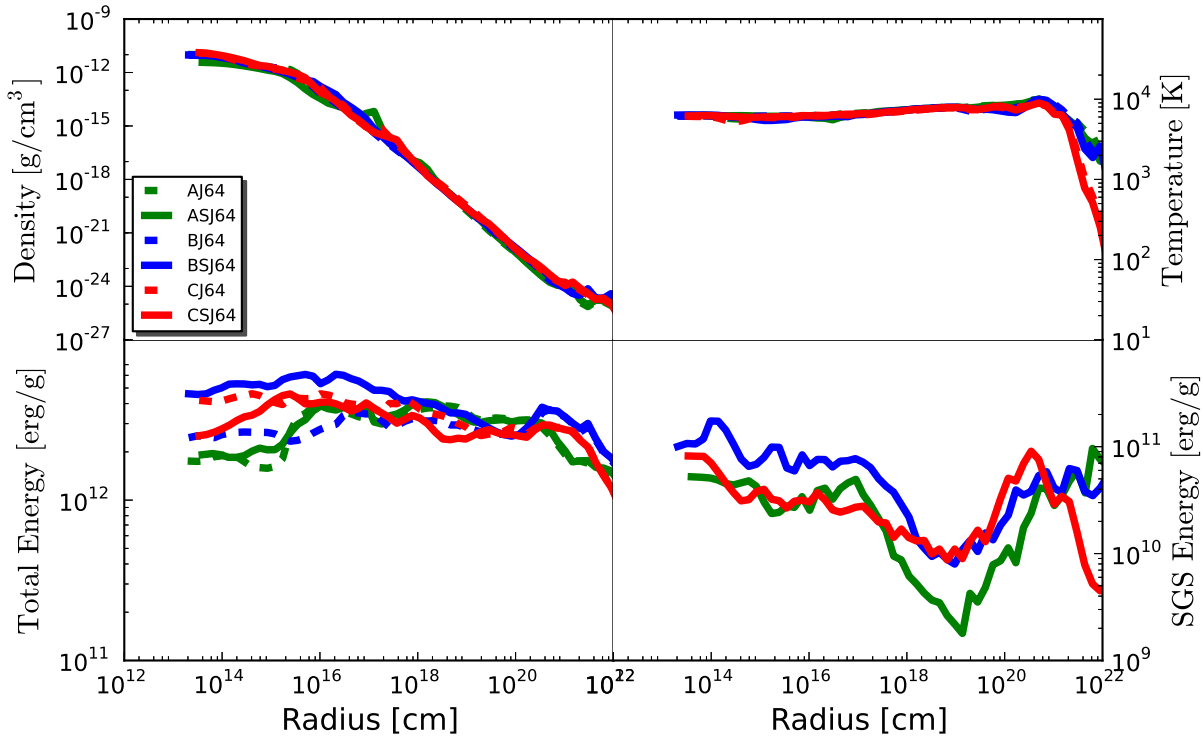}
\end{minipage}
\end{tabular}
\caption{The figure shows the radially binned spherically averaged radial profiles for halos A, B and C with and without SGS turbulence. The solid lines show the normal runs while the dashed lines represent the cases with SGS turbulence as shown in the legend. The upper left panel of the figure shows the density radial profiles. The temperature radial profiles are depicted in the upper right panel. The bottom left panel of the figure shows the total energy radial profiles. The averaged SGS energy radial profiles are depicted in the bottom right panel of the figure. }
\label{fig1}
\end{figure*}

\begin{figure*}
 \hspace{-4.0cm}
\centering
\begin{tabular}{c c}
\begin{minipage}{6cm}
\includegraphics[scale=0.4]{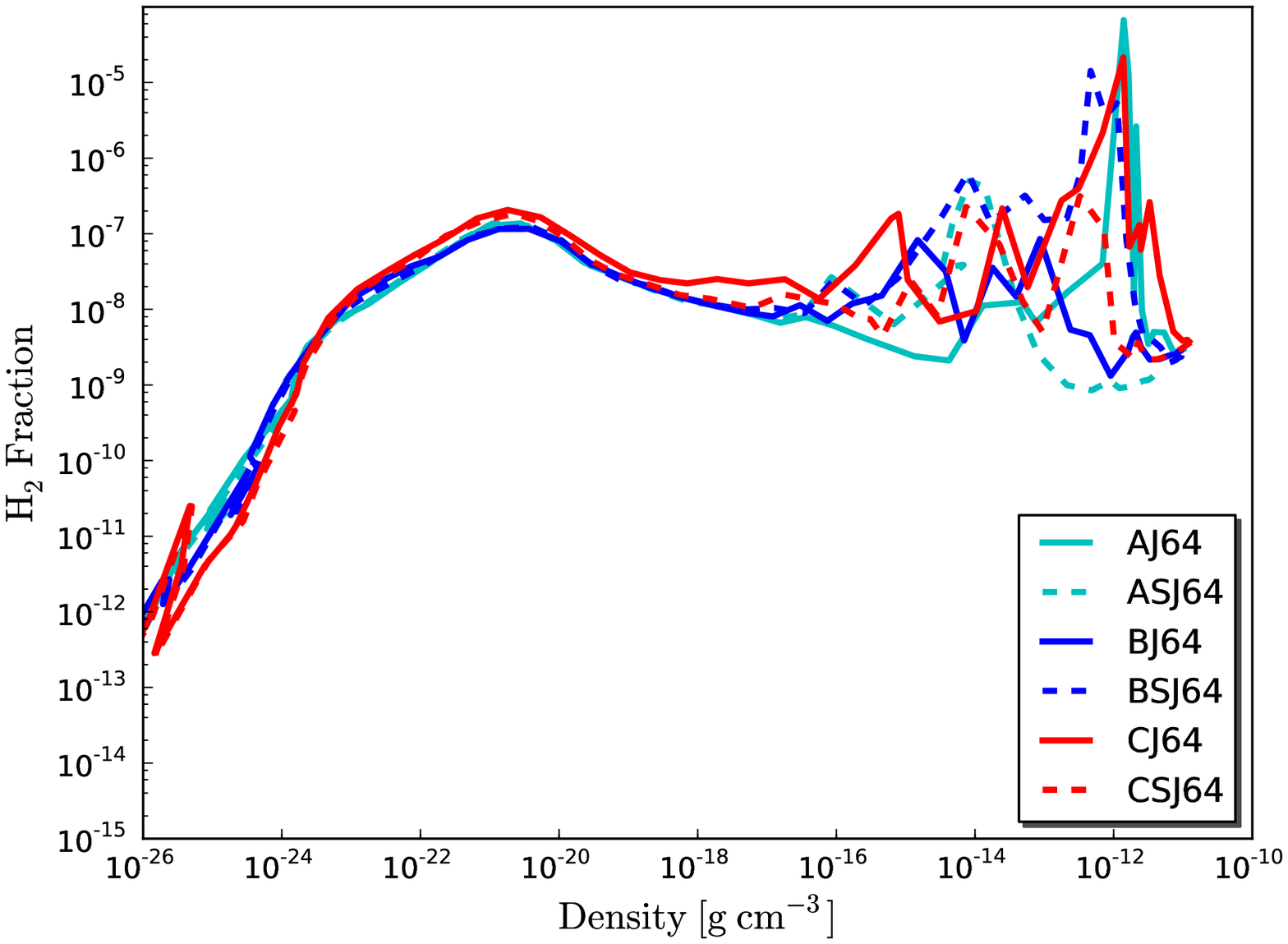}
\end{minipage} &
\hspace{2cm}
\begin{minipage}{6cm}
\includegraphics[scale=0.4]{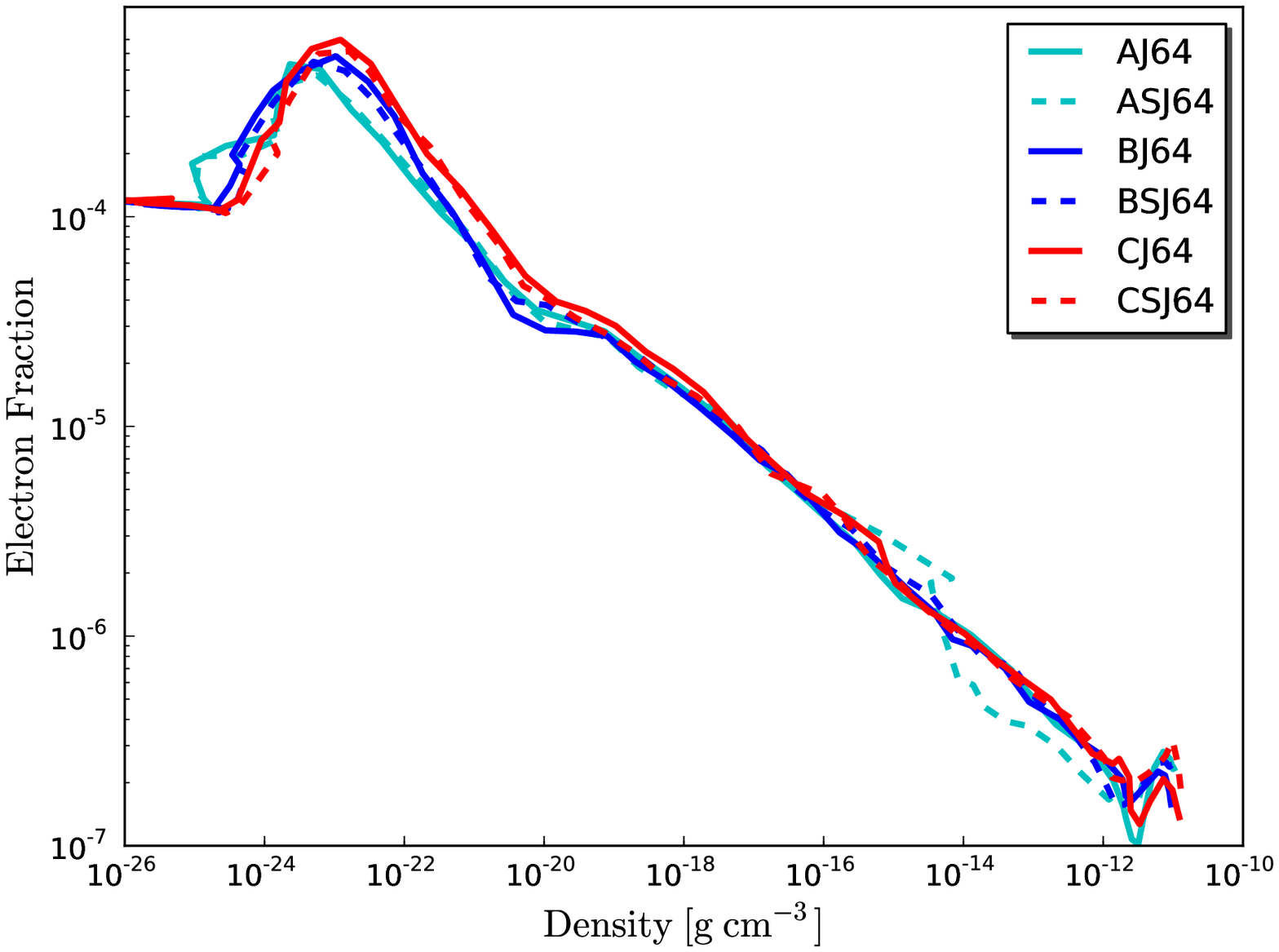}
\end{minipage}
\end{tabular}
\caption{The left panel of the figure shows the $\rm H_{2}$ abundance for three different haloes. The solid lines show the cases with no SGS turbulence and the dashed lines show the cases with SGS turbulence. The corresponding electron fraction is shown in the right panel.}
\label{fig3}
\end{figure*}

% \begin{figure*}
% \centering
% \begin{tabular}{c c}
% \begin{minipage}{6cm}
% \includegraphics[scale=0.5]{ABHJ640021_volume_render.pdf}
% \end{minipage} &
% \begin{minipage}{6cm}
% \includegraphics[scale=0.5]{ABHSJ640020_volume_render.pdf}
% \end{minipage} \\  \\
% 
% \begin{minipage}{6cm}
% \includegraphics[scale=0.5]{BBHJ640021_volume_render.pdf}
% \end{minipage} &
% 
% \begin{minipage}{6cm}
% \includegraphics[scale=0.5]{BBHSJ640021_volume_render.pdf}
% \end{minipage} \\ \\
% 
% \begin{minipage}{6cm}
% \includegraphics[scale=0.5]{BHJ640022_volume_render.pdf}
% \end{minipage} &
% 
% \begin{minipage}{6cm}
% \includegraphics[scale=0.5]{BHSJ640021_volume_render.pdf}
% \end{minipage} 
% \end{tabular}
% \caption{This figure shows the volume rendered images of the density for the central 100 AU region of the halo. The left panels of the figure show the density images for the halos without a sub grid scale turbulence from top to bottom halo A, B and C respectively. The density images for SGS turbulence corresponding to the left panels are shown in the right panel. }
% \label{fig4}
% \end{figure*}

\begin{figure*}
 \hspace{-4.0cm}
\centering
\begin{tabular}{c c}
\begin{minipage}{6cm}
\includegraphics[scale=0.4]{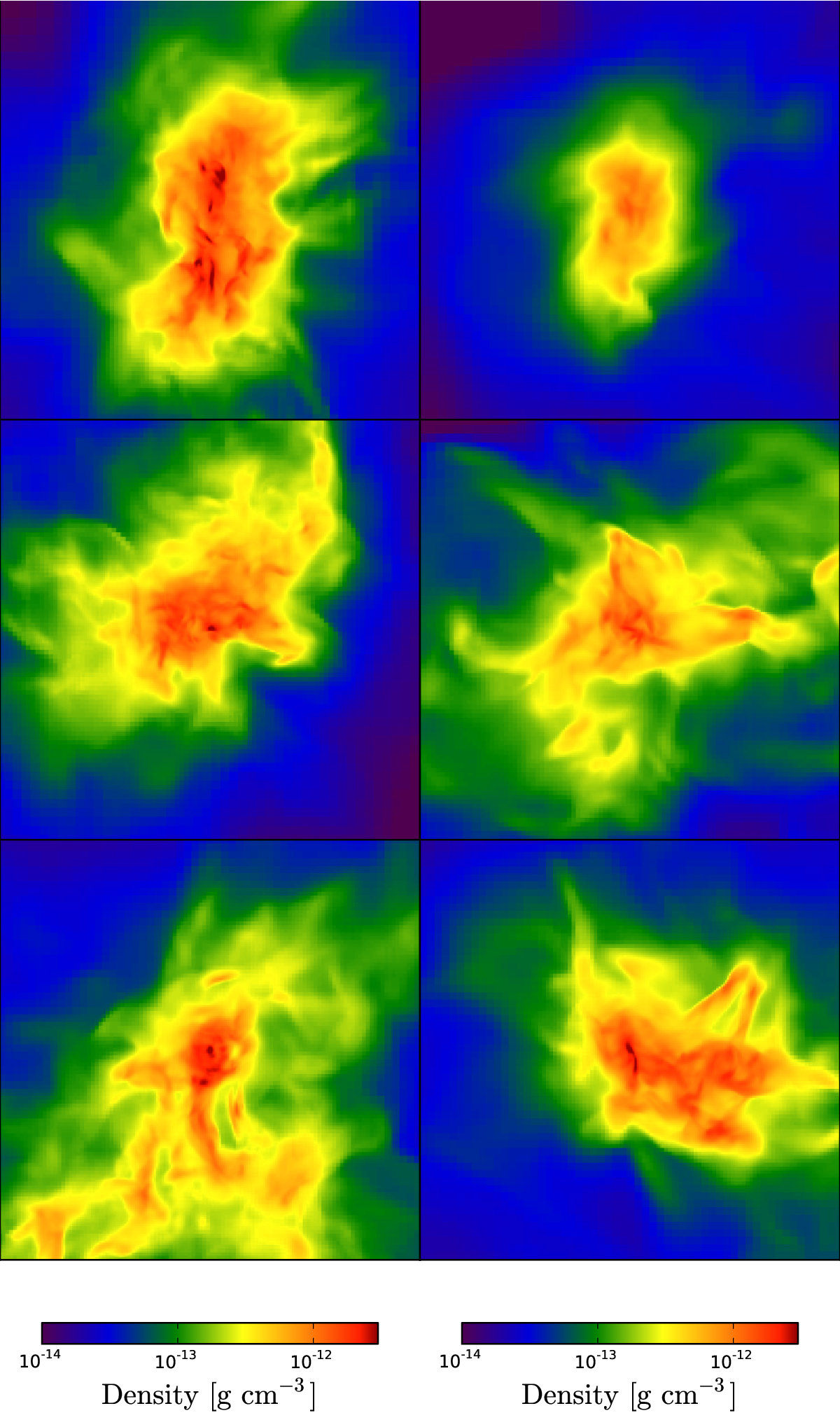}
\end{minipage} &
\hspace{2cm}
\begin{minipage}{6cm}
\includegraphics[scale=0.4]{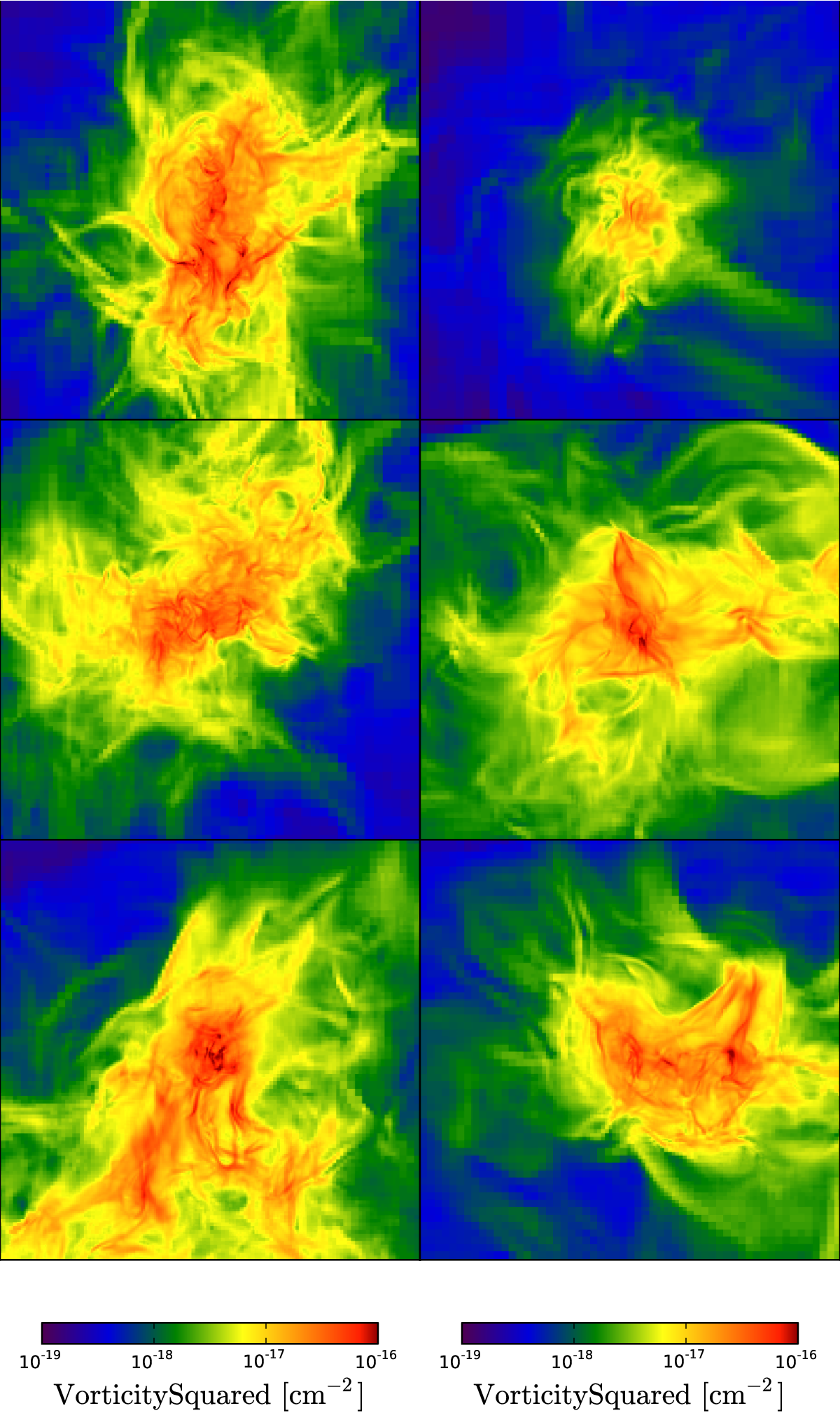}
\end{minipage}
\end{tabular}
\caption{The left panel of this figure shows the density projections for the central 500 AU region of the halo. The images are shown for three different halos A, B and C from top to bottom respectively. The left side of left panel shows the normal runs while the right side depicts the cases with SGS turbulence. The right panel of the figure shows the density weighted vorticity projections corresponding to the density projections in the left panel. It can be noticed that halos with SGS turbulence model have more compact structures.}
\label{fig5}
\end{figure*}

\begin{figure*}
\centering
\includegraphics[scale=0.8]{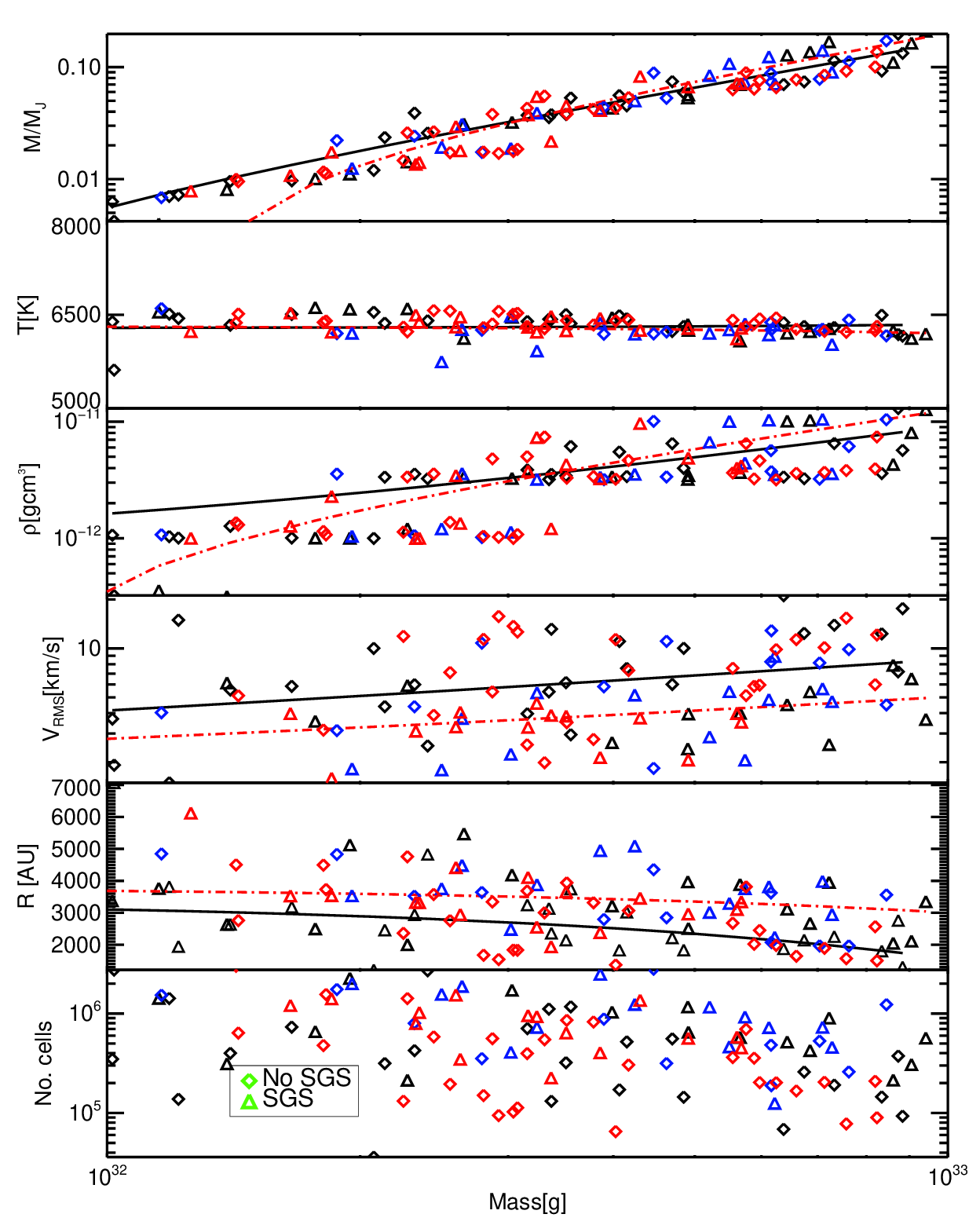}
\caption{The properties of the clumps are plotted against their masses in this figure. Diamonds show the data points for normal runs while data points for SGS turbulence cases are represented by triangles. Black, red and blue colors of the symbols represent the data for three different halos (i.e., A, B and C). The black solid lines are the fits to normal cases and the dot-dashed red lines are the fits to SGS cases. These clumps have sub-solar masses and are gravitationally unbound.}
\label{fig6}
\end{figure*}

\section{Results}

\subsection{Resolution comparison}

In this study, we have performed simulations resolving the Jeans length by 16, 32 and 64 cells (hereafter called $\rm J_{16},J_{32},J_{64}$ respectively) and compared the results with SGS turbulence model. The properties of the halo for different Jeans resolutions with and without the SGS turbulence model are shown in figure \ref{fig}. The results are examined for the same peak density (i.e., $\rm 10^{-11}~gcm^{-3}$). The averaged density radial profiles for all the cases are depicted in the top left panel of figure \ref{fig}. The density profiles show almost $\rm R^{-2}$ behavior according to the expectation of a isothermal collapse. The bumps in the density profile for $\rm J_{16}$ and $\rm J_{32}$ cases indicate the presence of under-resolved density clumps in the vicinity of the central object. The overall density profiles agree well for different resolutions.

The top right panel of figure \ref{fig} shows the averaged temperature radial profiles. It can be noticed that for all the cases the halo is heated up to its virial temperature and then cools by atomic line cooling. The temperature in the center of the halo remains about 7000 K. This shows that thermal properties of the halo are independent of resolution as well as SGS turbulence. The total energy for different Jeans resolutions and SGS turbulence cases is shown in the bottom left panel of figure \ref{fig}. The value of total energy increases at larger radii and becomes almost constant with a value of $\rm 3 \times 10^{12}~erg/g$ in the center of a halo. It is worth noting that these quantities are approximately converged, and do not depend strongly on the subgrid-scale model. Unlike in minihaloes, the results of such systems are thus likely more robust, as the thermal evolution is close to isothermal and less sensitive to minor changes in the dynamics.

% WS >>>
While the total energy is roughly the same for the three runs without SGS model, we see a non-monotonous behavior with resolution if the SGS model is applied. However, by considering the morphology of the halo in figure \ref{fig0}, it becomes clear that there are only little or no turbulent structure in the lower-resolution runs. For $\rm J_{16}$, turbulence is definitely under-resolved, while $\rm J_{32}$ could be an intermediate case. Consequently, the turbulent cascade cannot be sufficiently resolved in these runs. For the highest resolution case (i.e., $\rm J_{64}$), the structure inside the halo appears to be well resolved. The central region of a few hundred AU in size has become highly turbulent and clumpy. The turn-around in the trend for the total energy profiles indicates that Kolmogorov-like turbulence begins to be resolved for $\rm J_{64}$. This is plausible because the scale separation between the production of turbulence by gravity and grid scale is almost two decades, but the lower decade is 
strongly affected by numerical viscosity. 

Our results are also consistent with earlier studies \citep{2011ApJ...731...62F}.
The profiles of the SGS turbulence energy for the different resolutions are shown in the bottom right panel of figure \ref{fig}. Since the grid scale decreases as the refinement levels become higher, the expectation based on the Kolmogorov scaling law would be that the SGS turbulence energy decreases toward the center and with the maximal resolution. However, this applies only to homogeneous turbulence. Since turbulence production tends to be stronger in the center of the halo, the resolution-dependence is compensated and yields a nearly constant SGS energy in the central region. The comparable plateaus of the SGS energy for $\rm J_{32}$ and $\rm J_{64}$ also indicate that the enhancement of SGS turbulence production in the $\rm J_{64}$ case compensates the smaller grid scale in comparison to $\rm J_{32}$. The SGS energy has a peak around radii of $\rm 5 \times 10^{21}$ cm ($\rm 2~kpc$), which coincides with the drop in the temperature profiles. The outward decrease of the SGS energy at larger radii shows 
that turbulence is mainly produced as a result of the collapse, but not by hydrodynamical instabilities outside of the halo.

% The outward decrease of the SGS energy at large radii, on the other hand, shows that turbulence is mainly produced 
% as a result of the collapse, but not by hydrodynamical instabilities outside of the halo. Indeed, the radius at which the SGS energy drops
% coincides with the same feature in the temperature profiles. 
% WS: we should not explain away the differences by effects that could just as well affect the non-SGS runs ("different substructure,
% rotational velocity, ..."). Also note that radii cannot be identified with length scales.
%The SGS energy initially increases at larger radii because of turbulence cascade and gets almost saturated at smaller scales. The variation in %SGS energy comes from the intermittent flows occurring in the center of halo. The SGS energy is very small fraction of the total energy. It is %very low for SGS turbulence $\rm J_{32}$ case as most of the energy is retained in resolved scales.  
Furthermore, comparisons of the morphology of the halo for the three different Jeans resolutions reveal significant differences between the runs with and without SGS model (see figure \ref{fig0}). This is most obvious from the low-resolution runs that the SGS model is limited by the poorly resolved turbulence. However, we will concentrate on the highest-resolution case for quantitative comparisons in the next section.
% <<< WS

\subsection{Study of different Haloes}

\subsubsection{Global Dynamics of collapse}

We have performed six cosmological simulations for three different haloes (named A, B and C) with a constant strength $\rm J_{21}=10^{3}$ of the $\rm H_{2}$ photo-dissociating radiation field. The masses of the haloes and their collapse redshifts are listed in table \ref{table1}. The results obtained from cosmological simulations conducted in this work are presented in the following subsections.
The density fluctuations collapse under the gravitational instability as they decouple from the Hubble flow. These small fluctuations merge with each other to form larger haloes in accordance with the standard paradigm of structure formation. In the early phases of the collapse gas falls in the dark matter potentials and gets shock-heated during the nonlinear evolution phase. Gravitational energy of the halo is continuously transferred to kinetic energy of the gas and dark matter during the course of virialization.

Here we report the study of three haloes of different masses but with same Jeans resolution to compute the variation from halo to halo. Again, we compare our results with SGS turbulence model. The averaged density radial profiles for the haloes A, B and C with and without SGS model, centered at the densest cell are shown in the upper left panel of figure \ref{fig1}. The maximum density in our simulations is a few $\rm 10^{-11}~gcm^{-3}$. It can be seen from the figure that all cases follow almost $\rm R ^{-2}$ behavior which would be expected from a isothermal collapse. There is a bump in the density profile of halo A with SGS turbulence model which is an indication of fragmentation. In the very central region the density profile is flat which corresponds to the local Jeans length in all cases. These density profiles are comparable to the previous studies \citep{2002Sci...295...93A,2008ApJ...682..745W,2009Sci...325..601T,2012ApJ...745..154T}.

The average radial profile of total energy is shown in the bottom left panel of figure \ref{fig1}. The total energy of the system is a few times $\rm 10^{12}~erg~g^{-1}$ for all the cases. The figure shows the total energy for three different halos with and without SGS turbulence is converged. The increase in the total energy radial profile towards the smaller radii is due to gas infall in the center of halo.   

The specific subgrid scale energy for the three different halos is shown in the bottom left panel figure \ref{fig1}. At larger radii, the build-up of turbulent energy increases sharply because of the turbulence cascade and then gets saturated at smaller radii due to the enhanced turbulence dissipation rate. This evolution of SGS energy is according to the expectations of large eddy simulations \citep{2009ApJ...707...40M}. 

\begin{table}

\begin{center}
\caption{The halo masses and their collapse redshifts are listed in this table.}
\begin{tabular}{cccccc}
\hline
\hline

Model	& Mass			& Collapse redshift          \\

 & $\rm M_{{\odot}} $	& z   	      \\ 
\hline                                                           \\
 A	 & $\rm 8.06 \times 10^{6}$			& 11.9	      \\		
 B	  & $\rm 4.3 \times 10^{6}$			& 11.3	  	      \\
 C	  & $\rm 3.2 \times 10^{7}$			& 14.1	  	       \\	
 
\hline
\end{tabular}
\label{table1}
\end{center}

\end{table}

\subsubsection {Thermodynamics}

The thermal evolution of the gas for different halos with same Jeans resolution is shown in the right panel of figure \ref{fig1}. During the process of virialization, the gas is heated up to its virial temperature (i.e., $\rm \ge 10^{4}$K) and subsequently cools by Lyman alpha radiation. Consequently, gas collapses almost isothermally with temperatures around 8000 K. The temperature profiles for all the cases with and without SGS turbulence are similar. There is no significant turbulent heating for SGS turbulence cases as seen in \cite{2009ApJ...707...40M} due to highly efficient atomic line cooling at these temperatures. The dissimilarities in the temperature profiles at larger radii appear due to the difference in the halo masses. The ubiquity of intense Lyman Werner radiation photo-dissociates the molecular hydrogen and $\rm H_{2}$ cooling remains suppressed. Our results are in agreement with previous studies \citep{2003ApJ...596...34B,2008ApJ...682..745W,2011A&A...532A..66L,2012A&A...540A.101L} and according to the expectation of theoretical 
models.

The H$_{2}$ abundance is shown in the left panel of figure \ref{fig3}. It can be noticed that the H$_{2}$ fraction increases at lower densities due to the rise in electron abundance during the non-linear phase of the collapse. At intermediate densities, the H$_{2}$ abundance becomes constant as gas cools, recombines and remains neutral with a constant temperature around 8000 K. The presence of sharp spikes in the H$_{2}$ fraction is due to the shocks occurring at the central densities due to collisional dissociation. In general, the H$_{2}$ fraction is lower than the universal value (i.e., $\rm 10^{-3}$). Therefore, the contribution of H$_{2}$ cooling in the thermal evolution of the haloes studied here is negligible.
%  The main pathway to photo-dissociate molecular hydrogen is direct photo-dissociation of H$_{2}$ molecules by the Solomon process.

The electron abundance corresponding to the H$_{2}$ fraction is depicted in the right panel of fig \ref{fig3}. It can be seen that the electron abundance is correlated with the H$_{2}$ fraction. At densities above 1 $\rm cm^{-3}$, the electron fraction increases because of virialization shocks and then continues to decline as gas becomes neutral. Small wiggles in the central electron fraction are triggered by shocks.

\subsubsection {Halo structure}

The state of the simulations with and without the SGS turbulence at the collapse redshift is illustrated by the density projections in the left panel of figure \ref{fig5}. Significant changes in the morphology of haloes are found in the presence of SGS turbulence in all haloes (i.e., A, B and C). It can be noted that haloes are highly turbulent and clumpy in both cases. These effects were not seen in the earlier studies due to the poor Jeans resolution. Our results confirm that one needs to resolve  the Jeans length with  at least  32 cells or higher to capture turbulent velocity fluctuations \citep{2011ApJ...731...62F,2012ApJ...745..154T}. Overall, halos in simulations with SGS turbulence are more compact and denser than their counterparts. The latter is expected due to the presence of an additional viscosity term. As demonstrated above, these haloes have a similar thermal evolution but the changes in the morphology arise as a consequence of unresolved subgrid scale energy computed via the SGS 
turbulence model and show substantial variation from halo to halo. The SGS energy is about 10 \% of total energy budget. Its effect is particularly enhanced on small scales, yielding rather different morphologies in the presence of the subgrid-scale model. The structure of the halo ''A'' 
without SGS turbulence clearly shows that dense clumps are very well separated from each other and may lead to the formation of a binary in this case. The further evolution of the simulations becomes computationally very demanding as the Jeans mass keeps decreasing. Here, we stopped our simulations after reaching the maximum refinement level. We plan to explore the further evolution of at least one halo in a companion paper. 

To determine the presence of turbulent velocity fluctuations, we have computed the fluid vorticity (i.e., $\nabla \times v$). Density weighted projections of the vorticity squared centered at the densest point are depicted in the right panel of figure \ref{fig5}. In the center of the haloes, large regions with high values of vorticity indicate the ubiquity of high turbulent energy. It is also noted that high values of the vorticity are correlated with the dense regions of the haloes. These vorticity plots further suggest the absence of coherent structures. The amount of vorticity is higher  compared to the SGS turbulence cases because of higher turbulent dissipation rates in SGS turbulence cases.

\subsubsection{ Properties of clumps}

In order to quantify the properties of the clumps found during visual inspection, we have employed the clump finder of \citet{1994ApJ...428..693W}. The properties of the clumps with and without SGS turbulence model for three different haloes (A, B and C) with power law fits are shown in figure \ref{fig6}. The top panel of figure \ref{fig6} depicts that clumps are not gravitationally bound as their masses are smaller than the Jeans mass and the number of clumps is generally higher in no SGS cases. The ratio of clump mass to  Jeans mass shows a power law behavior (i.e., $\rm M/M_{J} \propto M^{1.3}$). It is interesting that similar trends have been found in different studies exploring clumps in molecular clouds \citep{2009MNRAS.398.1082B}. We find that in simulations with the SGS turbulence model, the clumps have slightly higher masses as shown by the fit (red line in figure \ref{fig6}) and the number of low mass clumps is reduced compared to the no SGS cases. This is likely an effect of the turbulent 
viscosity, which provides an additional diffusion mechanism that counteracts the formation of low-mass clumps. The thermal properties of the clumps are depicted in figure \ref{fig6}, showing that the clumps in  simulations with SGS turbulence have almost the same temperature as those in the standard setup. The density of the clumps plotted against mass shows a bimodal distribution with clumps sitting at higher densities following a linear relation with $\rm M$. The notable difference is that clumps in SGS turbulence case have higher densities and the power law sharply drops for lower densities. The latter suggests that only the high-mass clumps manage to form in the presence of an additional turbulent viscosity, providing a distribution with more massive clumps on average. %Our results suggest that clumps formed in SGS turbulence cases would become even more massive during the later stages of the collapse.

The velocity dispersion in the clumps increases with the mass and follows a $\rm M^{0.38}$ power law. Again, it is seen that clumps with SGS turbulence have lower values of dispersion velocity but roughly follow the same trend. The radii of the clumps comply a $\rm M^{0.8}$ growth and clumps with SGS turbulence model have larger radii in comparison with no SGS cases. In the last panel of figure \ref{fig6}, we show that these clumps are well resolved at least by $\rm 10^{5}$ cells.

\section{Discussion and Conclusions}

% WS >>>
We have conducted high resolution cosmological simulations using the AMR code Enzo for three different halos with $\rm T_{vir} \geq 10^{4}$K irradiated by a constant strength of photo-dissociating background UV flux. In one set of simulations, we used the subgrid scale (SGS) turbulence model proposed by \citet{SchmNie06b} and \citet{2011A&A...528A.106S} to compute the kinetic energy of numerically unresolved turbulence and the associated stresses on resolved length scales. For comparison, we run these simulations also without the SGS model. Since a high dynamical range is crucial, we use two initial nested grids and insert up to 27 additional refinement levels during the course of simulations, corresponding to an effective resolution down to sub AU scales. To investigate resolution effects, we applied refinement at 16, 32 and 64 cells per Jeans length. The results from three haloes with different masses were examined to study the variation from halo to halo for a fixed resolution. The main conclusions from this study are the following:

\begin{itemize}
\item The global properties of the halo, in particular the radial profiles, are converged and can be used as a robust input for direct collapse models.
  \item Turbulent structures are observed for a Jeans resolution of at least $\rm \geq 32$ cells.
 \item The morphology of the halo and its clump properties are strongly influenced by taking into account SGS turbulence and typically more compact.
% \item \ch{Subgrid scale turbulence leads to the formation of compact structures.}
 \item The clump properties (i.e., $\rm M/M_{J}$, velocity dispersion) show a power law behavior against clump masses.
% \item  \ch{The presence of turbulence has numerous potential implications for the formation of first objects like efficient amplification of magnetic fields, formation and stability of accretion disks.}  
   
\end{itemize}

The gas in the atomic cooling halos is heated up to its virial temperature where Lyman alpha cooling comes into play and cools the gas down to 8000 K. The presence of an intense Lyman Werner UV radiation field of $\rm 10^{3}$ in units of $\rm J_{21}$ photo-dissociates the $\rm H_{2}$ molecules via the Solomon process. We employed grid resolutions of $16$, $32$ and $64$ cells per Jeans length to explore the convergence of global properties as well as the local morphology. It is important to note that the radial profiles of density, temperature and total energy are approximately converged, implying more robust results than previously reported for minihalos \citep{2012ApJ...745..154T}. We attribute the latter to the thermodynamics of these halos, which are considerably more robust in the presence of strong H$_2$ photodissociation \citep{2010ApJ...712L..69S}. In this case, the temperature evolution remains very close to isothermal, making it rather insensitive to local changes in the dynamics. With such a fixed thermal pressure, the resulting evolution during the collapse is therefore considerably more robust. We note that SGS turbulence does not have a strong impact on thermodynamical properties, again as a result of efficient Lyman $\alpha$ cooling. The typical unresolved fraction of the turbulence energy is $\rm \simeq 10 \%$ of the total energy.

Our results demonstrate with the highest resolution simulations that atomic cooling halos become highly turbulent. We computed the evolution of three different halos and found that the radially averaged properties are very similar for three different halos, but the morphology of the haloes varies considerably. The intense vorticity inside the haloes demonstrates the absence of coherent structures in fully turbulent regions.
However, since the non-linear coupling between gravity and turbulence converts gravitational potential energy into kinetic energy, an inertial range of hydrodynamical turbulence can only exist on length scales smaller than the Jeans length. For this reason, a sufficient range of length scales smaller than the Jeans length must be resolved to observe a turbulent cascade with an inertial sub-range of the Kolmogorov type (for low compressibility) in simulations. This pushes numerical simulations to their limits because an extremely high dynamical range is required. In our simulations with 16, 32 and 64 cells per Jeans length, we find significant differences concerning the central morphologies and the amount of turbulent structures. The latter confirms that turbulence is only marginally resolved even at the highest resolution \citep[see also][]{2011ApJ...731...62F}. This is also reflected by the radial profiles of the turbulent energy in the simulations with SGS model, for which no clear convergence trend is 
found. To verify convergence, higher resolution $\rm \geq$ 64 cells per Jeans length should be performed in the future. 

Based on the notion of large eddy simulations, one would naively expect that the application of an SGS model reduces the range of length scales that has to be resolved. However, large eddy simulations are applicable only if the turbulent cascade is partially resolved, i.~e., at least by a decade in scale space. On top of that, numerical dissipation typically necessitates an additional decade. In simulations of turbulent collapsing halos, this corresponds to scale separation between the the Jeans length, at which energy is injected by gravity, and the grid scale. Apart from that, lack of local isotropy at low resolutions poses a problem because we use a constant coefficient for the production of SGS turbulence energy by shear. This problem could be addressed, for example, by a localized SGS model with varying coefficients \cite[see][]{SchmNie06c}. However, this method has not been applied yet in AMR simulation. 

Overall, it is important to note that, while a subgrid-scale model does not yield convergence at low resolution, it certainly does improve the solution once a sufficiently high resolution is reached, such that its central assumptions are fulfilled. In fact, it is clear that direct numerical simulations resolving the turbulence over a sufficient range of scales cannot be pursued in the near future. Obtaining robust astrophysical results therefore requires both high numerical resolution, as well as subgrid models to account for effects still below the grid scale. On the basis of the current simulations, we can already draw relevant conclusions about the effects of numerically unresolved turbulence for the Jeans resolution of 64 cells. By comparing the results with and without SGS cases, it was found that the morphology of the halos obtained in the simulations with SGS model is significantly different from their counterparts. In general, their  central gas distributions are denser and more compact compared 
to the none-SGS cases. The latter provides an indication that larger accretion rates onto the central object can be expected due to the additional turbulent viscosity. In one case, the density structure of the halo shows a bimodal distribution if no SGS model is applied, which might result in the formation of a binary. Such structural implications need to be addressed employing numerical methods like sink particles or a pressure floor. It is thus worth exploring whether statistical differences can also be obtained for gravitationally bound clumps, and how much their accretion is enhanced via turbulent viscosity.

We noticed that structures become compact in the presence of SGS turbulence. This may have important implications for the accretion of mass to the central object, potentially favoring the higher accretion rates. In our simulations, we observed the formation of turbulent structures for resolutions of $\rm \geq 32$ cells per Jeans length and no accretion disk at this stage of the collapse. We do not follow the evolution of these haloes for longer dynamical times and therefore cannot make statements about the final fate of these haloes. According to theoretical predictions, it is likely that formation of disk may take place. Numerical investigations of the direct collapse scenario thus need to employ a sufficiently high numerical resolution as well as a turbulence subgrid model to determine realistic accretion rates.

% <<< WS

% 
% 
% \begin{figure}[htb]
% \centering
% \includegraphics[scale=0.3]{ABHJ640021_vorticity3.pdf}
% \caption{Density projections}
% \label{fig:figure 3}
% \end{figure}

%ABHJ640021_Density3.pdf

\section*{Acknowledgments}
The simulations described in this work were performed using the Enzo code, developed by the Laboratory for Computational Astrophysics at the University of California in San Diego (http://lca.ucsd.edu). We thank Matt Turk, Robi Banerjee, John Wise and Enzo developers for helpful discussions. This work was supported from the SFB~963 (project A12) {\em Astrophysical Flow Instabilities and Turbulence}. We also acknowledge the funding support from German Science Foundation. DRGS thanks for funding from the Deutsche Forschungsgemeinschaft (DFG) in the Schwerpunktprogramm SPP 1573 “Physics of the Interstellar Medium” under grant SCHL 1964/1-1. The simulation results are analyzed using the visualization toolkit for astrophysical data YT \citep{2011ApJS..192....9T}.

\bibliography{biblio1.bib}

\end{document}